\documentclass[aps,floatfix,nofootinbib,preprint]{revtex4-1}

\usepackage{graphicx}
\usepackage{epic}
\usepackage{eepic}
\usepackage{latexsym}
\usepackage{amssymb,amsmath}

\newcommand{\eq}[1]{(\ref{#1})}
\newcommand{\be}{\begin{equation}}
\newcommand{\ee}{\end{equation}}
\newcommand{\bea}{\begin{eqnarray}}
\newcommand{\eea}{\end{eqnarray}}

\newcommand{\vs}[1]{\vspace{#1 mm}}
\newcommand{\hs}[1]{\hspace{#1 mm}}

\def\a{\alpha}
\def\b{\beta}
\def\cc{\gamma}

\def\d{\delta}

\def\e{\epsilon}

\def\f{\phi}
\def\fr{\frac}
\def\F{\Phi}

\def\l{\lambda}

\def\m{\mu}
\def\n{\nu}

\def\r{\rho}
\def\s{\sigma}

\def\Th{\Theta}

\def\O{\Omega}

\def\o{\omega}

\def\del{\partial}

\let\bm=\bibitem
\def\nn{\nonumber}

\begin{document}

\title{Schr\"{o}dinger from Wheeler-DeWitt: The Issues of Time and Inner Product in Canonical Quantum Gravity} 

\author{Ali Kaya}

\email[]{alikaya@tamu.edu}
\affiliation{\vs{3}Department of Physics and Astronomy, Texas A\&M University, College Station, TX 77843, USA \vs{10}}

\begin{abstract}

\vs{5}
	
The wave-function in quantum gravity is supposed to obey the Wheeler-DeWitt (WDW) equation, however there is neither a satisfactory probability interpretation nor a successful solution to the problem of time in the WDW framework. To gain some insight on these issues we compare quantization of ordinary systems, first in the usual way having the Schr\"{o}dinger equation and second by promoting them as parametrized theories by introducing embedding coordinate fields, which yields first class constraints and the WDW equation. We observe that the time evolution in the WDW framework can be described with respect to the embedding coordinates, where the WDW equation becomes Schr\"{o}dinger like, i.e. it involves first order timelike functional derivatives. Moreover, the  equivalence with the ordinary quantization procedure determines a suitable Hilbert space with a viable probability interpretation. We then apply the same construction to general relativity by adding  embedding fields without any prior coordinate choice.  The reparametrized general relativity  has two different types of diffeomorphism invariance, which arises from world-volume and target-space reparametrizations. As in the case of ordinary systems, the time evolution can be described with respect to the embedding fields and the WDW equation becomes Schr\"{o}dinger like; the construction is almost identical to an ordinary parametrized field theory in terms of time evolution and Hilbert space structure. However, this time, the constraint algebra enforces the wave-function to be in a subspace of states annihilated by an operator that can be identified as the Hamiltonian. The implications of these results for the canonical quantization program, and in particular for the minisuperspace quantum cosmology, are discussed.

\end{abstract}

\maketitle

\section{Introduction} 

Due to the underlying gauge invariance related to coordinate transformations, the canonical analysis of general relativity yields a weakly vanishing Hamiltonian\footnote{In this paper, we will neglect possible surface terms that may arise in the canonical analysis by assuming a compact space without boundary. These surface terms are obviously very important as they define conserved charges and even nontrivial boundary dynamics.}  that involves first class constraints. After canonical quantization, one then obtains the Wheeler-DeWitt (WDW) equation where the Hamiltonian constraint operator acting on the wave-function vanishes ${\cal H}\Psi=0$. As opposed to the usual Schr\"{o}dinger equation $i\hbar \del_t\psi=H\psi$, there is no explicit time derivative of the wave-function in the WDW equation and hence it is usually thought to give a timeless theory; a static formalism without any time evolution (see \cite{r} for an account of this). The WDW equation is extensively used in quantum cosmology, yet mostly in the Euclidean signature, see  the no-boundary \cite{qc1} and tunneling \cite{qc2} wave-function proposals of the initial state of the universe. Although these proposals seem to offer a smooth beginning for the universe, the Lorentzian formulation based on the Picard-Lefschetz theory gives a completely different picture, a beginning with unsuppressed fluctuations  \cite{qc3,qc4}, which shows the subtlety of the problem. 

Somehow related to the issue of time and the Euclidean formulation, one rarely finds in the literature a proper discussion of an inner product for the WDW states. Evidently, this is crucial for an appropriate physical interpretation of the quantum theory. Identifying the proper degrees of freedom is a particular difficulty in that context. Another reason is that the WDW equation is second order in partial/functional derivatives and thus the issue is similar to finding a positive-definite  invariant inner product for the Klein-Gordon fields; an old problem which can actually be solved using the theory of pseudo-Hermitian operators \cite{alimost1,alimost2}. On the other hand, we showed in \cite{ak1,ak2} that  the scale factor of the universe can be used as a perfectly well defined time variable in some simple minisuperspace  cosmological models. Such a model becomes ordinarily quantum mechanical after gauge fixing similar to a time dependent harmonic oscillator. The Hamiltonian is given by  the square root of a non-negative Hermitian operator, which is the only peculiarity that can easily be solved by the spectral theory. Moreover, the big-bang singularity is naturally resolved; one can rigorously  prove that the Hamiltonian degenerates to the zero operator as one approaches the big-bang. The  model can be viewed as a concrete realization of the zero energy universe hypothesis \cite{ze}.  Curiously, the initial state of the universe becomes completely arbitrary, which is the only price to be paid. 

The problem of time in quantum gravity is a deep one and there are various proposals to tackle it. For example, it is possible to  formulate quantum mechanics without time using the presymplectic geometry leading to the concept of quantum evolving constants  \cite{rov1,rov2} (see also \cite{rov3} for issues in this formalism and \cite{rov4}).  According to this approach time is not defined at the fundamental level and unitary evolution appears within a certain approximation. Similarly, \cite{page} suggests that the state of the universe must be stationary as the universe is a closed system and  the observed dynamical evolution emerges with respect to an internal clock reading, where the clock corresponds to  a subsystem of dynamical variables. On the other hand, in ordinary quantum mechanics with a Hamiltonian bounded from below no dynamical variable can play the role of time in the conditional probability interpretation \cite{uw}. Note that the issue of time and the probability interpretation of the wave-function are inherently related problems in quantum gravity, for a detailed account see \cite{kuchar}. It is also possible to use WKB expansion to restore time in quantum cosmology, for a recent review see  \cite{wkbp}. 

General relativity is a highly nonlinear theory and finding a representation of the underlying gauge symmetry is nontrivial since the algebra of constraints has field dependent "structure constants".  However, one can imagine simpler  parametrization invariant theories that also have first class constraints and vanishing Hamiltonians; from the point of view of dynamical evolution these are essentially the same with general relativity except their phase space has a much more plain structure. Specifically, one can reformulate ordinary systems as parametrized theories by introducing embedding coordinates as dynamical variables \cite{ik1,ik2}. As we will see,  this formulation is different from introducing an auxiliary metric to make the theory reparametrization invariant, in particular the original and the parametrized theories are actually identical. The quantization of the parametrized theory involves the WDW equation and comparing this to the original Schr\"{o}dinger formulation should give valuable insights. 

Not surprisingly, the quantization of an ordinary classical mechanical system and it's parametrized version turns out to be fully equivalent: The WDW equation that arises in the parametrized theory becomes the Schr\"{o}dinger equation where the  embedding time variable plays the role of time. Similarly, for a self-interacting scalar field in flat spacetime, the  evolution can be described with respect to embedding coordinates and  the WDW equation becomes  Schr\"{o}dinger like, i.e. it is first order in functional derivatives corresponding to timelike deformations.  The WDW formalism offers a broader description that allows propagation of states from an arbitrary initial to an arbitrary final spacelike surfaces and it becomes identical to the standard quantization  when the evolution is restricted to constant time slices in flat space. 

General relativity has already diffeomorphism invariance, but it is still possible to introduce embedding fields as dynamical variables \cite{ik2}. Now, \cite{ik2} uses Gaussian coordinates as an auxiliary foliation that completely fixes the diffeomorphism invariance and then introduces the embedding fields (see \cite{ikg1}  and \cite{ikg2} for generalizations). The new theory regains  diffeomorphism invariance corresponding to the changes of the embedding coordinates. Here, we  show that the embedding fields can be introduced without any prior gauge fixing and this actually doubles the diffeomorphism invariance, there are now  world-volume and target-space reparametrizations, where the embedding coordinates can be thought to define a map from a world-volume to a target-space.  The canonical analysis performed in \cite{ik2} actually deals with the world-volume gauge invariance. In any case, the reparametrized theory enlarged with the embedding coordinates is fully equivalent to the original general relativity. 

As we will discuss, the canonical structure of the reparametrized general relativity turns out to be almost identical to the parametrized scalar field; the time evolution can be described with respect to the embedding fields and the WDW equation becomes Schr\"{o}dinger like. There is no particular difficulty in introducing a Hilbert space with a proper probability interpretation. However, as an important distinguishing feature, the constraint algebra enforces the states to be constrained in a subspace annihilated by an operator that can be identified as the Hamiltonian. Thus, the time evolution is trivial, i.e. the states  do not evolve from one spacetime slice to another, yet the quantum mechanical probability interpretation is obvious. 

Finally in this paper, we apply the above construction to minisuperspace quantum cosmology of a real scalar field in order to clarify the Hilbert space structure and the physical interpretation of the wave-functions. The time evolution with respect to the embedding coordinate exactly takes the Schr\"{o}dinger form where the WDW operator plays the role of a Hamiltonian. Demanding this Hamiltonian to be Hermitian (with a suitable ordering) determines the inner product between the states and possible boundary conditions that must be imposed. Here, both the scale factor of the universe $a$ and the scalar field $\f$ are taken to be dynamical variables (and the embedding coordinate becomes time) as suggested by this construction. Like in the general case,  the physical states are confined in the zero energy subspace, but other than this peculiarity the system is  a typical quantum mechanical one. For the normalizability of physical states, and hence for the consistency of the minisuperspace model, it is crucial to have a discrete energy spectrum around zero as oppose to a continuous one. This is an important restriction, which is most of the time ignored in the literature. 

\section{Quantization of Parametrized  Theories} 

Let us first consider an ordinary mechanical system having the action 
\be\label{s1}
S=\int dt\left[\fr12 \dot{x}^2-V(x)\right],
\ee
which describes a particle moving in one-dimension in the potential $V(x)$, where  the dot denotes the $t$-derivative. To make the model time reparametrization invariant one can introduce a free parameter $\tau$, and take $t=t(\tau)$ and $x=x(\tau)$ as the dynamical variables. The new action can be written as  
\be\label{s2}
S=\int d\tau \,t' \left[\fr12 \left(\fr{x'}{t'}\right)^2-V(x)\right],
\ee
where the prime denotes $\tau$-derivative. Eq. \eq{s2} is invariant under $\tau$-reparametrizations that infinitesimally take the form 
\bea
&&\d t= \e\, t'\nn\\
&&\d x =\e \,x'
\eea
where $\e=\e(\tau)$. The reparametrized theory is indeed equivalent to \eq{s1}, which can be seen by gauge fixing $t=\tau$ in \eq{s2}. 

Note that this method of making the theory reparametrization invariant  is different than introducing an auxiliary metric on the world-line; while the latter would reduce the number of degrees of freedom, here \eq{s1} and \eq{s2} are identical. Note also that $t'$ can also be viewed as a world-line metric, but there is a difference between assuming a variable or its time derivative to be a fundamental field. From \eq{s2} one can see that the equation obtained from the variation of $t$ is identically satisfied once the equation of motion of  $x$  is assumed to hold.  This is a generic feature of this method, i.e. {\it the field equations of  the embedding coordinates are automatically satisfied when the equations of the other fields are assumed to hold.} 

Applying the canonical analysis to \eq{s2} gives an identically  vanishing Hamiltonian. There is a first class constraint which can be enforced by a Lagrange multiplier $\l$ and the action can be written as 
\be\label{ss1}
S=\int d\tau \left[P_t t' + P_x x' +\l\left(P_t+\fr12 P_x^2+V\right)\right].
\ee
One can choose the gauge $t=\tau$ and solve $P_t$ from the constraint to obtain
\be\label{ss2}
S=\int dt \left[P_x \dot{x} -\left(\fr12 P_x^2+V\right)\right].
\ee
This is the action \eq{s1} written in the Hamiltonian form, which again shows the classical equivalence of the two descriptions. 

In the quantum theory, let the ordinary and the parametrized theories have the wave-functions $\psi(t,x)$ and $\Psi(t,x)$, respectively. They obey $i\del_t\psi=H\psi$ and ${\cal H}\Psi=0$, where $H=\fr12 p_x^2+V$ and ${\cal H}=P_t+H$. Obviously, the Schr\"{o}dinger  and the WDW equations become identical. To achieve the full equivalence of the two descriptions the WDW inner product must be defined as $\left<\Psi_1|\Psi_2\right>=\int dx\, \Psi_1^*(t,x)\Psi_2(t,x)$, although \eq{ss1} would imply integrating over $t$ as well. Namely, without this comparison, one would naively interpret $|\Psi|^2$ as a probability distribution over $t$ and $x$. However, the equivalence requires that in the WDW formulation the time variable $t$ must be singled out as a non-dynamical parameter and the inner product must be introduced over the {\it real} degree of freedom $x$ (see \cite{nc} where this construction is generalized to non-commutative spacetime).

The above example is fairly simple but the situation becomes more intriguing in the field theory case. Consider a self interacting real scalar field in {\it flat space} whose action is 
\be\label{fs1} 
S=\int d^4X \left[-\fr12 \del_\m\phi\del^\m\f-V(\f)\right].
\ee
To obtain the parametrized version, one treats $X^\m=X^\m(y^\a)$ and $\f=\f(y^\a)$ as dynamical variables \cite{ik1}.  Geometrically, $X^\m(y^\a)$ corresponds to some foliation of the flat space, where $y^\a$ denotes the intrinsic coordinates. The new action can be written as
\be\label{fs2} 
S=\int d^4y \, \sqrt{-\cc}\left[-\fr12 \cc^{\a\b} \del_\a\phi\del_\b\f-V(\f)\right],
\ee
where $\cc_{\a\b}=\del_\a X^\m\del_\b X^\n\eta_{\m\n}$ is the induced metric on the foliation. The theory is invariant under coordinate changes $\tilde{y}^\a=\tilde{y}^\a(y^\b)$ where $X^\m$ and $\f$ transform like scalars. The infinitesimal transformation generated by the vector field $k^\a$ becomes
 \bea
 &&\d X^\m=k^\a\del_\a X^\m,\nn\\
 &&\d \f=k^\a \del_\a\f.\label{fsd}
 \eea
Eqs. \eq{fs1}  and \eq{fs2} describe the same dynamics, which can be seen by fixing the gauge freedom of coordinate transformations with the identity map $X^\m=\d^\m_\a y^\a$.   

For the canonical analysis, one may decompose $y^\a=(\tau,y^i)$, where $\tau$ is a time coordinate, and introduce  $P_\m=\d S/\d X'^\mu$ and $P_\f=\d S/\d \f'$ as conjugate momenta, where again the prime is the $\tau$-derivative. Not surprisingly, applying a Legendre transformation gives an identically  vanishing Hamiltonian, but there are four constraints which can be written as 
\bea
&&\del_i X^\m P_\m +\Phi_i=0,\label{momc0}\\
&&n^\m P_\m +\Phi=0.\label{hamc0} 
\eea
Here $n^\m$ is the future pointing unit normal to the constant $\tau$-surfaces obeying $n_\m\del_iX^\m=0$, $n_\m n^\n=-1$,  and 
\bea
&&\Phi_i=\del_i\f P_\f,\nn\\
&&\Phi=\fr{1}{2\sqrt{h}}P_\f^2+\fr12 \sqrt{h}\,h^{ij}\,\del_i\f\del_j\f+\sqrt{h}\,V(\f), \label{ffi}
\eea    
where $h_{ij}=\cc_{ij}$ is the induced spatial metric. The constraints can be enforced  by Lagrange multipliers $\l$ and $\l^i$ to obtain  
\be\label{fss1} 
S=\int d^4y \left[P_\m X'^\m+P_\f \f'+\l\left(n^\m P_\m +\Phi\right)+\l^i\left(\del_i X^\m P_\m +\Phi_i\right)\right].
\ee
Note that $n^\m$ is uniquely fixed from the spatial gradients  $\del_i X^\m$  and in particular it  does not depend on $X'^\m$, hence  \eq{fss1} is in the Hamiltonian form. At this point, one can deparameterize the theory by solving  $P_\m$ from the constraints and imposing the gauge $X^\m=\d^\m_\a y^\a$, which give 
\be\label{fss2} 
S=\int d^4X \left[P_\f \dot{\f}-\left(\fr{1}{2}P_\f^2+\fr12\,\d^{ij}\,\del_i\f\del_j\f+\,V(\f)\right)\right],
\ee
where $X^\m=(t,x^i)$ and the dot, as above, denotes the $t$-derivative. This shows the classical equivalence of the original and the parametrized theories since \eq{fss2} is the action \eq{fs2} written in the Hamiltonian form.  

In the Schr\"{o}dinger picture, the quantization of the original non-parametrized theory can be achieved by imposing the canonical commutation relation $[\f(x),P_\f(\tilde{x})]=i\d^3(x-\tilde{x})$. The wave function is a functional $\psi=\psi[\f;t]$  obeying the  Schr\"{o}dinger equation
\be\label{sche}
i\frac{\del}{\del t} \psi =\int d^3 x \left(\fr{1}{2}P_\f^2+\fr12\,\d^{ij}\,\del_i\f\del_j\f+\,V(\f)\right)\psi
\ee
where one can use  $P_\f(x)=-i\d/\d\f(x)$. The inner product can formally be introduced by a path integral  so that
\be\label{ip0}
\left<\psi|\o\right>=\int D\f\, \psi[\f]^*\,\o[\f].
\ee
Of course, these formal definitions are most of the time ill defined and one needs renormalization theory to make sense of them.  Moreover, unlike quantum mechanical systems with finite degrees of freedom, the representation of the canonical commutation relations is by no means unique. In any case, one may assume that the quantization is well understood, at least in the perturbation theory around Fock space of free particle states. 

Let us now discuss the WDW quantization of the parametrized theory \eq{fss1}. The quantization is based on the configuration space variables $X^\m(y^i)$ and $\f(y^i)$, which can be thought to live on an initial  $\tau=\tau_0$ surface. The wave-function is a functional $\Psi[X^\m(y^i),\f(y^i)]$ obeying the WDW equations 
\bea
&&\left(\del_i X^\m P_\m +\Phi_i\right)\Psi=0,\label{momc}\\
&&\left(n^\m P_\m +\Phi\right)\Psi=0,\label{hamc} 
\eea
where $P_\m=-i\d/\d X^\m(y)$, $P_\f=-i\d/\d \f(y)$ and $\Phi$, $\Phi_i$ are given in \eq{ffi}. These are functional differential equations imposed  at each spatial position $y^i$ and unlike the  Schr\"{o}dinger equation \eq{sche} there is no $\tau$-derivative of the wave-function hence they do not describe an {\it explicit} time evolution of the state. Recall that $n^\m$ denotes the unit normal vector field which is uniquely determined by $\del_i X^\m$. 

\begin{figure}
	\centerline{\includegraphics[width=9cm]{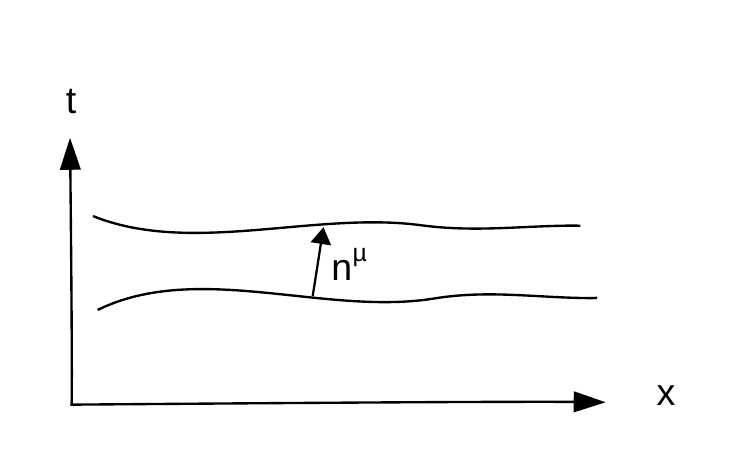}}
	\caption{An arbitrary initial value surface and it's deformation along the normal vector.} 
	\label{fig1}
\end{figure}

The constraints are similar to those of general relativity; while \eq{momc} can be viewed as the momentum constraint, \eq{hamc} becomes the Hamiltonian constraint. Indeed, \eq{momc} can be seen to imply (assuming the suitable ordering where the momenta is placed to the right) 
\be\label{spd}
\Psi\left[X^\m(y^i),\f(y^i)\right]=\Psi\left[X^\m(y^i+k^i),\f(y^i+k^i)\right],
\ee
where $k^i=k^i(y)$ are arbitrary (infinitesimal) functions. Therefore,  \eq{momc} ensures the invariance of the wave-function under spatial diffeomorphisms where $X^\m$ and $\f$ transform like scalars. On the other hand, \eq{hamc} determines the change in the wave-function when the embedding fields are deformed  along a direction perpendicular to the initial $\tau=\tau_0$ surface, i.e.  given an arbitrary (infinitesimal)  function $\e=\e(y^i)$, the shifted wave-function  $\Psi\left[X^\m(y^i)+\e \,n^\m,\f(y^i)\right]$  is determined from $\Psi\left[X^\m(y^i),\f(y^i)\right]$ to first order in $\e$. Namely, one has 
\bea
\Psi\left[X^\m(y^i)+\e \,n^\m,\f(y^i)\right]&&\simeq\Psi\left[X^\m(y^i),\f(y^i)\right]+\int d^3 y\, \e\, n^\m\, \fr{\d}{\d X^\m(y)}\Psi\left[X^\m(y^i),\f(y^i)\right]\nn\\
&&=\Psi\left[X^\m(y^i),\f(y^i)\right]-i\int d^3 y\, \e\, \Phi\,\Psi\left[X^\m(y^i),\f(y^i)\right]\label{expansion} 
\eea
where in the first line the first argument  is expanded in $\e$ and in the second line \eq{hamc} is used. This  describes  how the wave-function changes from an initial surface to its deformation along the normal vector, which can clearly be viewed as {\it  time evolution} (see Fig \ref{fig1}).   

Finally, an inner product should be introduced to complete the quantization.  As suggested by the simple mechanical system discussed above, one should define the inner product by integrating out the $\f$ field leaving the embedding coordinates intact, i.e. 
\be\label{ip2} 
\left<\Psi|\Omega\right>=\int D\f\, \Psi^*[X^\m(y),\f]\,\Omega[X^\m(y),\f],
\ee
where $D\f=\prod_{y^i} d\f(y^i)$. The path integral measure is invariant\footnote{The Jacobian of the linear transformation $\f\to\f+k^i\del_i\f$ is indeed equal to 1.} under the change of variable $\f\to\f+k^i\del_i\f$, as a result  the momentum constraint \eq{momc} ensures that the inner product is invariant under spatial diffeomorphisms
\bea
\left<\Psi|\Omega\right>&=&\int D\f\, \Psi^*[X^\m+k^i\del_i X^\m,\f]\,\Omega[X^\m+k^i\del_iX^\m,\f]\nn\\
&=&\int D\f\, \Psi^*[X^\m,\f-k^i\del_i \f]\,\Omega[X^\m,\f-k^i\del_i\f]\nn\\
&=&\int D\f\, \Psi^*[X^\m,\f]\,\Omega[X^\m,\f].\label{ipsd} 
\eea
Similarly, the  Hamiltonian constraint \eq{hamc} guarantees the invariance of the inner product under {\it time translations}  
\bea
\left<\Psi|\Omega\right>&=&\int D\f\, \Psi^*[X^\m+\e n^\m,\f]\,\Omega[X^\m+\e n^\m,\f]\nn\\
&=&\int D\f\, \Psi^*[X^\m,\f]\,\Omega[X^\m,\f].\label{iptt} 
\eea
which can be seen using  \eq{ffi} in \eq{hamc} and applying functional integration by parts. 

Note that the WDW equation describing the timelike deformations \eq{hamc} involves a {\it first order} timelike functional derivative. Therefore, it is Schr\"{o}dinger like and the structure of the quantum theory is not much different than an ordinary quantum mechanical system.  

It could be interesting to elaborate more on the WDW quantization of the parametrized scalar field theory (see e.g. \cite{mad1,mad2} for possible issues that may arise). However, our main interest in this paper is to understand how time evolution emerges in the WDW formalism and for that one can actually consider a much simpler  construction  by only adding a single timelike embedding coordinate $t(\tau,y^i)$ to \eq{fs1} (thus, one does not need to worry about the algebra of constraints and can avoid possible complications like the existence of anomalies). In that case, the parametrized action \eq{fs2} becomes (one now has $X^i=y^i$) 
\be\label{tact}
S=\int d^3 y\, d\tau \, t'\, \left[\fr12\left(1-\del_i t\del_i t\right)\fr{\f'^2}{t'^2}+ \del_i t \del_i\f\fr{\f'}{t'} -\fr12 \del_i\f\del_i\f-V\right],
\ee
which is invariant under local time reparametrizations generated by 
 \bea
&&\d t=k \,t',\nn\\
&&\d \f=k \,\f' ,\label{fst}
\eea
where $k=k^\tau(\tau,y^i)$ is an arbitrary function. This is the only local symmetry and  the action \eq{tact} is no longer invariant under spatial diffeomorphisms. As before the Hamiltonian identically vanishes and there is a first class constraint 
\be\label{nhamc}
(1-\del_i t\del_i t)^{1/2}P_t-n^i \del_i\f P_\f + \Phi=0,
\ee
where 
\be\label{f24}
\Phi=\fr{1}{2} n^t P_\f^2+\fr12  (1-\del_i t\del_i t)^{1/2}\left[\d^{ij}+n^i n^j\right] \del_i\f\del_j\f+(1-\del_i t\del_i t)^{1/2}V
\ee
and the unit normal vector field of the foliation $t=t(\tau,y^i)$ is given by
\be
n^t=\fr{1}{(1-\del_j t \del_j t)^{1/2}},\hs{5}n^i=\fr{\del_i t}{(1-\del_j t \del_j t)^{1/2}}.
\ee
Note that $n^\m$ does not depend on $t'$ as pointed out above.  It is also possible to obtain \eq{nhamc} from \eq{fss1} after imposing the gauge $X^i=y^i$ and solving the corresponding momenta $P_i$ from the momentum constraint \eq{momc0}. 

In the Hamiltonian formulation, the phase space variables  $t(y^i)$, $P_t(y^i)$, $\f(y^i)$  and $P_\f(y^i)$ are defined on an initial $\tau=\tau_0$ surface and the states in the quantum theory become functionals  $\Psi=\Psi[t(y^i),\f(y^i)]$ annihilated by \eq{nhamc}
\be\label{wdw0}
\left[(1-\del_i t\del_i t)^{1/2}P_t-n^i \del_i\f P_\f + \Phi\right]\Psi=0,
\ee
where the ordering ambiguity in the first term is solved in an obvious way.  The inner product can be defined as in \eq{ip2}  by a path integral over $\f$
\be\label{ip3} 
\left<\Psi|\Omega\right>=\int D\f\, \Psi^*[t(y^i),\f]\,\Omega[t(y^i),\f].
\ee
Although  \eq{ip3} seems to depend on $t(y^i)$, the WDW equation \eq{wdw0} ensures this is not the case, i.e.  $\d/\d t(y^i)$ applied to \eq{ip3} actually vanishes.\footnote{From \eq{wdw0}, the action of  $\d/\d t(y^i)$  on a state equals the sum of two terms involving $n^i \del_i\f P_\f$  and  $\Phi$. One can see using functional integration by parts  that $\Phi$ is Hermitian, $\Phi^\dagger=\Phi$ under the inner product \eq{ip3} hence its contribution vanishes. On the other hand, the action of $n^i\del_i P_\f$ vanishes  thanks to the invariance of the path integral measure under $\phi\to\f+n^i\del_i\f$.}  Thus, the inner product does not depend on the slice  $t=t(y^i)$ and in particular it does not change under time evolution corresponding to the deformations of the surface $t=t(y^i)$. 

Clearly, the WDW formalism offers a more general description than the usual quantization since it allows one to specify an initial state over an arbitrary spacelike surface and it describes time evolution to an arbitrary final spacelike surface. As expected, one can  recover the usual quantization  when the evolution along constant time slices in flat space $t(y)=t$ are considered. Defining  $\psi[t,\f(y)]=\Psi[t,\f(y)]$ (i.e. $\psi$ is obtained by replacing $t(y)=t$ in $\Psi$),  \eq{wdw0} yields
\be
i\frac{\del}{\del t} \psi = \int d^3 y \, \F(y)\, \psi,
\ee 
which is exactly the Schr\"{o}dinger equation \eq{sche} of the original theory.\footnote{Note that given a functional $F[t(y)]$ and a function $f(t)$, which is obtained from $F[t(y)]$ by setting $t(y)=t$, one has $df/dt=\int dy \, \d F/\d t(y)$. Also for $t(y^i)=t$ the normal vector field becomes $n^t=1$, $n^i=0$.}  Furthermore, the inner product \eq{ip3} becomes identical to \eq{ip0}, showing that WDW formalism recovers the usual quantization.  
 
These elementary examples give valuable information about the WDW equation as they show how time evolution emerges relative to the embedding coordinates instead of a fixed background time. They also show, by comparison to the original quantization procedure, how an inner product must be introduced in the WDW formalism. 

\section{Reparametrized General Relativity} 

In this section we would like to apply the findings of the previous section to pure general relativity. It is possible to enlarge the field space of general relativity so that the field variables become the metric $g_{\m\n}(y^\a)$ and the embedding coordinates $X^{\m}(y^\a)$.  Here $X^\m$ denotes some coordinates on the spacetime manifold $M$ and $X^\m(y^\a)$ represents an embedding map from an identical manifold\footnote{Actually, $M$ and $\tilde{M}$ do not have to be identical since all our considerations will be local. This condition is necessary for the equivalence of the original and reparametrized theories.} $\tilde{M}$ that has coordinates $y^\a$. In the language of string theory, $M$ is the target space and $\tilde{M}$ is the world-volume.  The embedding map is assumed to preserve the pseudo-Riemannian  structure and the enlarged theory is defined on $\tilde{M}$, i.e. both $g_{\m\n}$ and $X^{\m}$ must be seen as functions of $y^\a$. 

The reparametrized general relativity has actually been considered by Isham and Kuchar in \cite{ik2}. However, Isham and Kuchar  use Gaussian coordinates that fixes the diffeomorphism invariance of the original theory completely. They  then introduce the embedding coordinates as new variables and study the algebra of constraints corresponding to the diffeomorphisms of the embeddings. As we will discuss in a moment, the reparametrized general relativity has actually two different types of diffeomorphism symmetry that must be  distinguished from each other and the analysis of \cite{ik2} actually deals with one of them. 

The action of the reparametrized theory involving  $g_{\m\n}$ and $X^\m$ can be written as
\be\label{pga} 
S=\fr12 \int d^4y \,\sqrt{-\cc}\, R(\cc)
\ee
where 
\be\label{cc}
\cc_{\a\b}=\del_\a X^\m \del_\b X^\n g_{\m\n},
\ee
$\cc=\det \cc_{\a\b}$ and $R(\cc)$ is the Ricci scalar of $\cc_{\a\b}$.  Formally, \eq{pga} is like rewriting the Einstein-Hilbert action  under a  coordinate change  $X^\m\to y^\a$ and it might be tempting  to revert back to the original action  $S=\int d^4 X\, \sqrt{-g} \,R(g)$, which naively makes the construction trivial. However, both the integral measure $d^4 X$ and the derivatives like $\del g_{\m\n}/\del X^\r$ are composite objects in the parametrized  theory which must be expressed in terms of the presumed fundamental variables; for instance  $d^4 X=\det( \del_\a X^\m)\, d^4 y$.  

Varying the action \eq{pga} with respect to $g_{\m\n}$ and $X^\m$ implies 
\bea
&&G^{\a\b}\del_\a X^\m \del_\b X^\n=0,\nn\\
&&\del_\a \left(\sqrt{-\cc}\, G^{\a\b} \del_\b X^\n g_{\m\n}\right) =0,
\eea
where $G^{\a\b}$ is the Einstein tensor.  Clearly the second equation is automatically satisfied once the first one is assumed to hold. As noted above, this is a generic feature of this formalism which is actually a manifestation of the equivalence of the original and parametrized theories. Given any solution of the original theory  $g_{\m\n}(X)$ having $G_{\m\n}(X)=0$, the parametrized theory has a solution with $X^\m=X^\m(y^\a)$ and $g_{\m\n}=g_{\m\n}(X^\r(y^\a))$, where $X^\m(y^\a)$ is an arbitrary set of embedding fields. 

Eq. \eq{pga} is invariant under two different diffeomorphisms.  The first is related to coordinate changes in $\tilde{M}$, which is infinitesimally generated by a (world-volume) vector $k^\a(y^\a)$ 
\bea
&&\d X^\m=k^\a\del_\a X^\m,\nn\\
&&\d g_{\m\n}=k^\a \del_\a g_{\m\n}.\label{wvd}
\eea
Note that  $X^\m$ and $g_{\m\n}$ are treated like (world-volume) scalars. One can see from the definition \eq{cc} that the variations \eq{wvd} induce $\d \cc_{\a\b}={\cal L}_k \cc_{\a\b}=k^\cc \del_\cc \cc_{\a\b}+\del_\a k^\cc  \cc_{\cc\b}+\del_\b k^\cc \cc_{\cc\a}$, where ${\cal L}_k$ is the Lie derivative.  The second symmetry is related to the coordinate changes in $M$ generated by a spacetime vector $l^\m(y^\a)$ so that
\bea
&&\d X^\m=l^\m,\nn\\
&&\d g_{\m\n}=-\del_\a l^\r X_\m^\a  g_{\r\n}  -\del_\a l^\r X_\n^\a  g_{\r\m},    \label{tsd}
\eea
where $X_\m^\a$ is the inverse of $\del_a X^\m$, i.e. 
\be\label{invx}
X_\m^\a \del_\a X^\n = \d^\n_\m, \hs{5} X_\m^\a \del_\b X^\m = \d^\a_\b.
\ee
One can see that $\d \cc_{\a\b}=0$ under \eq{wvd} and thus \eq{pga} is left invariant. Note also that \eq{tsd} acts nonlinearly on $X^\m$ as a shift symmetry. In the work of \cite{ik2},  the coordinate invariance in the target space has been fixed from the beginning, hence \eq{tsd} is lost. 

It is not difficult to see that \eq{pga} is fully equivalent to general relativity. One can use the gauge invariance   to set $X^\m=\d^\m_\a y^\a$, which eliminates  $X^\m$ from the dynamical system. This then yields the usual Einstein-Hilbert action
\be\label{eha}
S=\fr12 \int d^4 X\, \sqrt{-g} \,R(g).
\ee
The residual symmetries respecting the gauge choice $X^\m=\d^\m_\a y^\a$  must obey $k^\a \d^\m_\a = -l^\m$. The transformations \eq{wvd}  and \eq{tsd} then imply 
\be
\d g_{\m\n}=-l^\r\del_\r g_{\m\n} -\del_\m l^\r  g_{\r\n}  -\del_\n l^\r  g_{\r\m}=-{\cal L}_l g_{\m\n},   
\ee
which is the usual infinitesimal diffeomorphism map of general relativity. 

\section{Canonical Analysis} 

Let us now study the Hamiltonian formulation of \eq{pga}. As we have pointed out, this analysis generalizes that of  \cite{ik2} because \cite{ik2}  starts with a gauge fixed theory  in Gaussian coordinates and we are not employing any prior gauge fixing. The induced metric \eq{cc} can be decomposed into the  Arnowitt-Deser-Misner (ADM) components by defining
\bea
&&h_{ij}=\del_i X^\m \del_j X^\n g_{\m\n},\nn\\
&&N_i=\del_i X^\m X'^{\n}g_{\m\n},\label{nni}\\
&&N=-n_\m X'^\m,\nn
\eea
where as before  $y^\a=(\tau,y^i)$, the prime denotes  $\tau$-derivative and $n_\m$ is the future pointing unit normal vector of the foliation $X^\m(y^\a)$ which is uniquely fixed by $\del_i X^\m$ and $g_{\m\n}$ so that
\be\label{dnv}
n_\m \del_i X^\m=0,\hs{5} n_\m n_\n g^{\m\n}=-1.
\ee
In particular it is crucial to note that $n_\m$ does not depend on $X'^\m$. From the definitions  \eq{nni} and \eq{dnv} it is possible to see that\footnote{The dependence of $n_\m$ on $X'^\n$ in \eq{n} is misleading as one can directly check  from \eq{n} that $(\del n_\m/\del X'^\n)n^\m=0$ and $(\del n_\m/\del X'^\n)\del_i X^\m=0$, showing $\del n_\m/\del X'^\n=0$.}  
\bea
&&N^2=N^i N_i - X'^\m X'^\n g_{\m\n},\nn\\
&&n_\m=\fr{1}{N}(X'^\n -N^i \del_iX^\n) g_{\m\n}, \label{n}\\
&&h^{ij}\del_i X^\m \del_j X^\n=g^{\m\n}+n^\m n^\n,
\eea
where $N^i=h^{ij}N_j$ and $h^{ij}$ is the inverse of $h_{ij}$.  

Using the above equations, the reparametrized action \eq{pga} can be expressed (after some integration by parts)  in the ADM form
\be\label{adma}
S=\fr12 \int d\tau\, d^3 y \, \sqrt{h} \left[N R^{(3)}+\fr{1}{N}\left(K_{ij}K^{ij}-K^2\right)\right],
\ee
where $D_i$ is the derivative operator and  $R^{(3)}$ is the Ricci scalar of $h_{ij}$, $K_{ij}=\fr12(h'_{ij}-D_i N_j-D_j N_i)$,  $K=h^{ij}K_{ij}$, $h=\det h_{ij}$ and all indices are manipulated by $h_{ij}$. Note that our definition of  $K_{ij}$ differs by a factor of $1/N$ compared to the standard expression of the extrinsic curvature. 

From \eq{adma} the canonical momenta conjugate to $g_{\m\n}$ and $X^\m$ can be  found as
\bea
&&P^{\m\n}=\fr{\sqrt{h}}{2N}\left(K^{ij}-Kh^{ij}\right)\del_i X^\m \del_j X^\n,\nn\\
&&P_\m=-\fr12 \sqrt{h} \, R^{(3)} \,n_\m+ \fr{\sqrt{h}}{2N^2}\left(K^{ij}K_{ij}-K^2\right)n_\m-\fr{\sqrt{h}}{N}\left(K^{ij}-Kh^{ij}\right)D_i(g_{\m\n}\del_jX^\n). \label{gxmom}
\eea
Applying the Legendre transformation to obtain the Hamiltonian gives
\be
\int d\tau d^3 y\left[ P^{\m\n}g'_{\m\n}+P_\m X'^\m-{\cal L} \right]=0,
\ee
where the Lagrangian density ${\cal L}$ can be read from \eq{adma}. Not surprisingly,  the Hamiltonian of the parametrized theory vanishes {\it identically}. This should be contrasted with the standard canonical analysis of general relativity where the Hamiltonian vanishes as a result of the field equations of the lapse and the shift, which become Lagrange multipliers. 

From \eq{gxmom} one can see that there are 8 primary constraints which can be written as 
\bea
&&P^{\m\n}n_\n=0,\nn\\
&&\del_i X^\m P_\m+\Phi_i=0,\label{rgrc}\\
&&n^\m P_\m +\Phi+2P^{ij}D_i(\del_j X^\n g_{\m\n})n^\m=0,\nn
\eea
where 
\bea
&&\Phi_i=2P^{jk}D_k(\del_j X^\n g_{\m\n})\del_i X^\m,\nn\\
&&\Phi=\fr{2}{\sqrt{h}}\left[P^{\m\n}P_{\m\n}-\fr12 P^2\right]-\fr12 \sqrt{h}R^{(3)}. \label{nc}
\eea
Here the  $(i,j,k...)$  and $(\m,\n,\r...)$  indices are manipulated by $h_{ij}$ and $g_{\m\n}$, respectively. We also define 
\bea
&&P=P^{\m\n}g_{\m\n},\nn\\
&&P_{ij}=\del_i X^\m \del_j X^\n P_{\m\n}.\label{pij}
\eea
Note the the covariant derivative in the last term in \eq{rgrc} can be replaced by ordinary partial derivative since the terms involving the Christoffel symbols vanish once they are contracted with $n^\m$.  Using the definition \eq{pij} and the inverse of $\del_\a X^\m$ \eq{invx}, which can be determined as  
\bea
&&X_\m^i=g_{\m\n}h^{ij}\del_j X^\n+\fr{N^i}{N}n_\m,\nn\\
&&X_\m^\tau=-\fr{1}{N}n_\m,
\eea
one can see that $\sqrt{h}(K^{ij}-Kh^{ij})/2N=P^{ij}$. It is possible to express  the last 4 constraints in \eq{rgrc} in a more compact form as 
\be\label{thm}
{ \Theta}_\m=P_\m+\fr12 \sqrt{h} \, R^{(3)} \,n_\m -\fr{2}{\sqrt{h}}\left[P^{\m\n}P_{\m\n}-\fr12 P^2\right]n_\m+2 P^{ij}D_i\left(\del_j X^\n g_{\m\n}\right)=0,
\ee
which actually follows from the definition of momentum $P_\m$ given in \eq{gxmom}. The action governing the dynamics can be written as
\be\label{hahf}
S=\int d\tau d^3 y \left[P^{\m\n}g'_{\m\n} +P_\m X'^\m + \b_\m P^{\m\n}n_\n +\l(n_\m P^\m +\Phi) + \l^i(\del_i X^\m P_\m+\Phi_i)\right]
\ee
where $\b_\m$, $\l$ and $\l^i$ are Lagrange multipliers that enforce the constraints\footnote{As we will see below, there are also secondary constraints enforced by the constraint algebra. These turn out to be the first class constraints of the gauge fixed action \eq{hamacgr} and in principle they can be added  to the action \eq{hahf} with Lagrange multipliers.}  \eq{rgrc}. 

At this point, one may like to establish the equivalence of the reparametrized theory and  general relativity at the Hamiltonian level. Imposing the gauge\footnote{In general, one can set $X^\m=X^\m(y^\a)$ for some arbitrary embedding functions, and solve for $P_\m X'^\m$ to obtain the reduced Hamiltonian \eq{hahf}.} $X^\m=\d^\m_\a y^\a$ and solving the corresponding momenta $P_\m$ gives the action
\be\label{hamacgr} 
S=\int dt \, d^3 X\left[P^{ij}\dot{h}_{ij}+\fr12 N\sqrt{h} R^{(3)}-\fr{2N}{\sqrt{h}}\left(P^{ij}P_{ij}-\fr12 P^2\right)-2P^{ij}D_i N_j\right] 
\ee
where  $P=P^{ij}h_{ij}$ and we have used $P^{\m\n}n_\n=0$ to eliminate the corresponding momentum components. Here  the spatial metric $h_{ij}$, the lapse $N$ and the shift $N_i$, which are originally defined in \eq{nni} as composite fields, become  fundamental degrees of freedom after gauge fixing. Eq. \eq{hamacgr} is the standard general relativity action written in the Hamiltonian form, which shows again the equivalence of the reparametrized and original theories.  

Other than the existence of an additional one, i.e. the first line in \eq{rgrc}, the structure of the constraints in the  reparametrized general relativity is almost identical to the parametrized scalar field theory, compare \eq{rgrc} to \eq{momc} and  \eq{hamc}. Especially, the constraints are  {\it linear} in the embedding momenta and  once  the time evolution is identified with respect to the  embedding coordinates, the WDW equation that follows from \eq{rgrc} becomes Schr\"{o}dinger like with a first order timelike functional derivative. Thus, the main obstacle in introducing a Hilbert space with an invariant inner product that has the standard quantum mechanical probability interpretation is resolved.\footnote{Obviously, technical challenges related to the gauge invariance and possible ordering ambiguities still exist, nevertheless, in principle, an invariant inner product  can be introduced as in the case of a scalar field \eq{ip2}. See \cite{woodard} that discusses gauge-fixing of a canonical inner product in the standard WDW quantization of gravity.} 

In this paper we are not going to study the constraint algebra of the parametrized general relativity. The fundamental Poisson brackets between canonically conjugate variables can be written as 
\bea
&&\left\{X^\m(y_1),P_\n(y_2)\right\}=\d^3(y_1-y_2),\nn\\
&&\left\{g_{\m\n}(y_1),P^{\r\s}(y_2)\right\}=\fr12\left(\d^\r_\m \d^\s_\n+\d^\s_\m\d^\r_\n\right)    \d^3(y_1-y_2).
\eea
In general, modifying the standard algebra of constraints in one way or another changes the spacetime structure and general covariance \cite{boj1}, which may have important  implications, see e.g. \cite{boj2} on how the modifications implied by loop quantum cosmology affect the no boundary proposal of \cite{hh}. In our case, we are not directly or indirectly employ any modification of the constraint algebra, here we simply rewrite the Hamiltonian formulation of general relativity that keeps the spacetime covariance manifest.  Indeed, one would expect $k^\a\del_\a X^\m \Th_\m$ and $l^\m\Th_\m$ to be the generators of the transformations \eq{wvd} and \eq{tsd} in the Hamiltonian framework. 

Obviously, general relativity must have peculiarities compared to an ordinary scalar field theory. Indeed, the additional constraint in \eq{rgrc}, which does not show up in the field theory, has important implications. To see this we calculate the following Poisson bracket 
\be\label{ecgr}
\left\{P^{\m\r}n_\r(y_1),\Theta_\n(y_2)\right\},
\ee
which must weakly vanish as a consistency requirement (in quantum theory the physical states are annihilated by the constraints, and thus their commutators must also annihilate the states). After a relatively long computation we obtain
\bea
&&n_\m (y_1)\left\{P^{\m\r}n_\r(y_1),\Theta_\n(y_2)\right\}n^\n(y_2)=-\fr12\, \Phi \, \d^3(y_1-y_2)+...\nn\\
&&\del_i X^\s g_{\s\m} (y_1)\left\{P^{\m\r}n_\r(y_1),\Theta_\n(y_2)\right\}n^\n(y_2)=\, \sqrt{h} D_j \left(\fr{1}{\sqrt{h}} P^j{}_i\right)\d^3(y_1-y_2)+...\label{sc}
\eea
where $\Phi$ is given in \eq{nc} and the dotted terms are multiplied by the primary constraint $P^{\m\n} n_\n$ and hence they weakly vanish. Other contractions of the free indices in \eq{ecgr} by $n^\m$ and $\del_iX^\m$ identically vanish. 

Eq.\eq{sc} implies two secondary constraints and the whole set can be written as  
 \bea
 &&P^{\m\n}n_\n=0,\nn\\
 &&\del_i X^\m P_\m+\Phi_i=0,\nn\\
 &&n^\m P_\m +2P^{ij}D_i(\del_j X^\n g_{\m\n})n^\m=0,\label{rgrc2}\\
 &&D_j \left(\fr{1}{\sqrt{h}} P^j{}_i\right)=0,\nn\\
  &&\Phi=0.\nn
 \eea
Note that $\Phi$ can be viewed as the Hamiltonian because it is quadratic in the momenta and it was determining the dynamical change in the wave-function under a timelike deformation in \eq{rgrc}. After setting $\Phi=0$,  the Hamiltonian and momentum constraints (the second and the third lines in \eq{rgrc2}) become structurally identical 
\bea
&&\del_i X^\m P_\m+2\del_i X^\m D_k(\del_j X^\n g_{\m\n})P^{jk}=0,\nn\\
&&n^\m P_\m +\del_i X^\m \del_j X^\n (\d_n g_{\m\n})\, P^{ij}=0,\label{rgrc3}
\eea
where $\d_n g_{\m\n}$ is the change in the metric given in \eq{tsd} with $l^\m=n^\m$. These can be made more manifest by using $P^{\m\n}n_\m=0$ in \eq{rgrc3}, which yields  
\bea
&&\del_i X^\m P_\m+(\del_i g_{\m\n}) P^{\m\n}=0,\nn\\
&&n^\m P_\m + (\d_n g_{\m\n})\, P^{\m\n}=0.\label{rgrc4}
\eea
In the quantum theory, these ensure the invariance of the wave-function $\Psi[X^\m,g_{\m\n}]$ under spatial  and timelike deformations which are given by \eq{wvd} and \eq{tsd}, respectively. The momentum and Hamiltonian constraints are now on an equal footing; they simply fulfill purely kinematical conditions arising from the covariance.  

The last two constraints in \eq{rgrc2} are similar to the original general relativity and of course it is quite hard to solve the corresponding WDW equations in the quantum theory. However, this is now merely a technical complication; the conceptual problem of time  is solved by referring to the embedding coordinates and there is no difficulty in introducing an inner product between states with an appropriate probability interpretation. The only peculiarity compared to the scalar field theory case is that  the states are now constrained in the zero energy subspace hence they do not dynamically evolve in time (they simply change according to the diffeomorphism invariance implied in \eq{rgrc4}). In the next section, we will quantify these findings in more detail in the simplified minisuperspace setup. 

\section{Minisuperspace Quantum Cosmology} 

Minisuperspace quantum cosmology is a consistent truncation of general relativity; most of the time it is simple enough to tackle analytically, yet it is an important testing framework to examine the fundamental issues in quantum gravity. It has been extensively studied in the literature for a long time, but there is no consensus on how to define the Hilbert space of states and how to interpret the time evolution.  In this section, we apply the construction studied above to shed some light on these issues. 
 
The standard minisuperspace considers a real self-interacting scalar field $\f$ on the spacetime
\be
ds^2=-N^2d t ^2+a^2d\vec{x}^2.\label{m1}
\ee
For purely time dependent fields $\f=\f(t)$, $a=a(t)$ and $N=N(t)$, the Einstein-Hilbert action \eq{eha} reduces to 
\be\label{ms1}
S=\int\, dt\, a^3\left[-\fr{3}{N}\fr{\dot{a}^2}{a^2}+\fr{1}{2N}\dot{\f}^2-NV(\f)\right],
\ee
where $V(\f)$ is the scalar potential and we set $M_p^2\equiv1/(8\pi G)=1$. Here, one treats $N$, $a$ and $\f$ as the main independent degrees of freedom. 

Let us  introduce the embedding time coordinate  as a new variable $t(\tau)$ along the lines detailed above and consider the action 
\be\label{ms2}
S=\int\, d\tau\, a^3\left[-\fr{3}{Nt'}\fr{a'^2}{a^2}+\fr{1}{2Nt'}\f'^2-Nt'V\right],
\ee
which is invariant under $\tau$-reparametrizations whose infinitesimal version generated by  $\e=\e(\tau)$ can be written as
\bea
&&\d t=\e\, t',\nn\\
&&\d N=\e\,  N',\nn\\
&&\d a=\e\, a',\label{mr1}\\ 
&&\d \f=\e\, \f'.\nn
\eea
The reparametrized action \eq{ms2} has another local symmetry generated by the variations 
\bea
&&\d t=\tilde{\e}\nn\\
&&\d N=-\fr{\tilde{\e}'}{t'}\,  N,\nn\\
&&\d a=0,\label{mr2}\\ 
&&\d \f=0,\nn
\eea
where $\tilde{\e}=\tilde{\e}(\tau)$. Eqs. \eq{mr1} and \eq{mr2} correspond to \eq{wvd} and \eq{tsd}, respectively. The reparametrized and the original theories are identical, which can be seen after gauge fixing $t=\tau$ in \eq{ms2} that yields \eq{ms1}. 

Applying the standard Legendre transformation for the canonical analysis shows that the Hamiltonian {\it identically} vanishes. The primary constraints that follow from the action \eq{ms2} become
\bea
&&P_N=0,\nn\\
&&P_t + N {\cal H}=0,\label{mspc}
\eea
where 
\be\label{nf}
{\cal H}=-\fr{1}{12a}P_a^2+\fr{1}{2a^3}P_\f^2+a^3V. 
\ee
These can be imposed by Lagrange multipliers so that the action can be written as  
\be\label{msh1}
S=\int\, d\tau\, \left[P_t t'+P_a a'+P_\f \f'+P_N N'+\b P_N+ \l\left(P_t + N {\cal H}\right)  \right].
\ee
The constraints and the action can be compared with the general expressions \eq{rgrc} and \eq{hahf}.

The Hamiltonian structure is almost identical to the ordinary mechanical system \eq{ss1}, except there is now an extra constraint $P_N=0$. In any case, this strong similarity suggests that the wave-function must be taken in the form $\Psi(a,\f;t)$.  The WDW equation corresponding to \eq{mspc} turns into Schr\"{o}dinger type with respect to the embedding time
\be
i\frac{\del}{\del t} \Psi = N {\cal H} \Psi,
\ee
where one should  treat $t$ as a parameter and $a$, $\f$ as dynamical variables.  One can then introduce the usual inner product 
\be\label{ip} 
\left<\Psi|\O\right>=\int_{-\infty}^{+\infty} \int_0^\infty w(a)\, da\, d\f \, \Psi^*\O
\ee
with a  possible wight function $w(a)$. The inner product is positive-definite and it becomes invariant assuming that ${\cal H}$ is a Hermitian operator ${\cal H}^\dagger ={\cal H}$. In that case, one has the standard  interpretation, i.e.  $|\Psi|^2\, w(a)\, da \,d\f$ gives a probability distribution over $a$ and $\f$.  If one places the momenta to the right of the coordinates, which seems to be a natural ordering implied by the momentum constraint like in \eq{spd}, the choice 
\be
w(a)=a,\hs{5}\Psi(0,\f)=0
\ee
and suitable falloff conditions as $\f\to\pm\infty$ ensures ${\cal H}^\dagger ={\cal H}$ (and there always exists  a self-adjoint extension since ${\cal H}$ is real).

Without the extra constraint $P_N=0$, the above construction pretty much corresponds to a standard quantum mechanical one (the variable $N$ can be removed by introducing the proper time $d\tau=Ndt$). The constancy of the constraint $P_N=0$ implies ${\cal H}=0$, or in the quantum theory the condition $P_N\Psi=0$ requires
\be
{\cal H}\Psi=0.\label{wdwo}
\ee
This is not a fancy condition, it only enforces the states to be confined in the zero-energy subspace in the Hilbert space. Consequently, the time evolution of the physical states is trivial. Nevertheless, all other standard quantum mechanical rules  like calculating probabilities or expectation values of an observable still apply. Even the measurement problem is technically the same, i.e. a state can collapse to another state as a result of a measurement at a given time where both states are in the zero-energy subspace (of course the prominent question of how to interpret a measurement on the universe is still remains). 

Note that \eq{wdwo} is the original WDW equation but it does not uniquely fixes  ${\cal H}$ since one can multiply it by an overall factor. Moreover, it is not obvious from \eq{wdwo} whether $a$ and $\f$ must be taken as dynamical variables or one of them maybe playing the role of time, and how an inner product must be introduced. Indeed one tends to interpret either $a$ or $\f$ as a time variable  in \eq{wdwo} (for a recent account see \cite{yeni2}), however defining a time independent positive-definite inner product  becomes a severe problem for either choice because of the quadratic momentum dependence. Unfortunately this issue, which is similar to finding a positive-definite invariant  inner product for the Klein-Gordon equation,  is  usually ignored in the literature. A direct comparison leads one inevitably to third quantization in quantum cosmology since the Klein-Gordon equation implies the second quantization.  

Also, imposing $P_N=0$ as a constraint is necessary if one would like to deal with the canonical quantization of general relativity, otherwise  the theory becomes modified  (see \cite{yeni1} whose approach actually implying a modification). It is possible to consider such generalizations like the unimodular gravity where one imposes $\det g_{\m\n}=-1$, see  \cite{uw2,uw}, but this corresponds to a modification of general relativity as discussed in \cite{uw2,uw}. 

A final technical remark regarding the normalization of the physical states deserves to be mentioned. Since ${\cal H}$ has a hyperbolic character, one would expect to find an unbounded energy spectrum. If this spectrum turns out to be continuous around the origin, then the zero-energy subspace becomes improper, i.e. it is not possible to normalize the states satisfying \eq{wdwo}  properly. This is a crucial technical obstacle for the canonical quantization program, which has also been emphasized in \cite{uw,uw2}. 

\section{Conclusions} 

Our motivation in this work was to gain some insight on the  issues about the  WDW equation by comparing the quantization of ordinary systems with their parametrized versions. This comparison showed us how to describe the time evolution and introduce a positive-definite invariant inner product in the WDW framework. We then applied the same construction to general relativity by enlarging the field space with the embedding variables. We have achieved this without a prior coordinate choice. Remarkably, the canonical structure of the reparametrized general relativity becomes almost identical to the parametrized scalar field theory. In particular, the WDW equation turns into  Schr\"{o}dinger  type, i.e. it now involves a first order functional time derivative that allows one to introduce an invariant inner product straightforwardly. Moreover, the time evolution can be described with respect to the embedding variables in a natural geometric way. 

The canonical structure of the reparametrized general relativity is covariant  from the target spacetime point of view and one only needs to give up the world-volume covariance by choosing a time coordinate.  This makes the construction much more obvious geometrically, as a result the time evolution in the spacetime can be described covariantly. Unlike the ordinary general relativity, the Hamiltonian vanishes identically which is a common property of manifestly reparametrization invariant theories. To appreciate the advantage of this formulation, one can for example think about quantum electrodynamics; although the physics does not depend on the gauge choice, there is a difference in clarity for many results obtained in the Lorentz vs. Coulomb gauges. 

The way the time evolution is described in the present work is different than the method of deparametrization where the Hamiltonian constraint is solved to for one variable that becomes an emergent time parameter and the dynamics of the remaining variables are governed by a new non-vanishing Hamiltonian, see e.g. \cite{ak1} where the scale factor of the universe becomes a time parameter in a minisuperspace cosmological model and \cite{son} where an extra scalar field becomes an emergent time parameter and the quantization of the remaining degrees of freedom is achieved in the Loop Quantum Gravity framework. Here, we also enlarge the field space by introducing the embedding coordinates but the theory is not deparametrized, e.g. the Hamiltonian still vanishes, yet the time evolution can still be described by referring to the embedding coordinates. 

There is an inherent relationship between the embedding coordinates and the spacetime metric that is worth commenting here. To be mathematically more precise, we actually extend the original configuration space of the theory by adding the space of all {\it Lorentzian foliations} (i.e. the foliations having 3-dimensional spacelike slices).  Let us call the configuration space of all Lorentzian foliations ${\cal M}_X$. Obviously, the metric  intrinsically plays a role in the definition of ${\cal M}_X$. If the embedding coordinates are used to parametrize ${\cal M}_X$, then they should obey conditions like $g_{\m\n}X'^\m X'^\n<0$ and $\det \,( g_{\m\n}\del_i X^\m \del_j X^\n)>0$. These conditions define ${\cal M}_X$ as an {\it open region} in the function space of all embeddings $X^\m(\tau,y)$ (since the conditions involve strict inequalities). When the background metric is fixed, as in the case of a scalar field in flat spacetime, ${\cal M}_X$ has a fixed structure.  On the other hand, if ${\cal M}_g$ denotes the space of all Lorentzian metrics defined on a given manifold, the configuration space of the extended general relativity becomes a warped product ${\cal M}_g\ltimes {\cal M}_X$, where  ${\cal M}_X$ has a nontrivial fibration over ${\cal M}_g$. In any case, since ${\cal M}_X$ is open, i.e. we assume that the infinitesimal variations of a Lorentzian foliation is also a Lorentzian foliation,  these considerations do not affect the local analysis we have carried out above. Moreover,  we do not need a detailed  knowledge of the global structure of ${\cal M}_X$, which can be quite difficult to determine, since one does not (functionally) integrate over the embedding coordinates in the quantum theory, e.g. in defining the inner product. As a result, although in the beginning we treat the embedding coordinates as dynamical variables, their eventual physical role becomes quite different than, say,  a usual scalar field. Namely, they do {\it not} have any dynamical evolution (note that the constraints including the Hamiltonian are linear in their momenta) and they  simply become the parameters describing the time evolution in the theory.  

As mentioned above, the similarity between the reparametrized general relativity  and the parametrized scalar field theory is eminent. Yet, general relativity has its own peculiarity i.e. the physical states are confined in the zero-energy subspace in the Hilbert space.  This might have important consequences for some minisuperspace cosmological models since the normalizability of states can be problematic depending on the energy spectrum. Moreover, the interpretation of the wave-function as a probability distribution over the scale factor $a$ and the scalar field $\f$ calls for a careful evaluation of some of the results. For instance, most of the time a path integral over the field variables is thought to describe a time  evolution which is misleading as we have discussed here. Besides the path integral measure is inherently related to  the inner product in the Hilbert space and in particular it depends on the weight function possibly appearing in \eq{ip}. All of the states in the  zero-energy subspace are physically acceptable and there is no apparent principle of selecting one in the theory, which shows the severeness of the "initial condition" problem in this context, similar to the cases discussed in \cite{ak1, ak2}. Obviously, there are still many issues to be addressed about the WDW quantization of gravity, nonetheless we believe the present formulation sheds some light on a few highly debated crucial problems.

\end{document}